\documentclass[12pt]{article}
\usepackage{amsfonts, amsmath, amssymb, amsthm}
\usepackage{color}
\usepackage{dsfont}
\usepackage[T1]{fontenc}
\usepackage{graphicx}
\usepackage{mathrsfs}
\usepackage{url}

\newtheorem{theorem}{Theorem}[section]
\newtheorem{proposition}[theorem]{Proposition}

\newtheorem{lemma}[theorem]{Lemma}

\theoremstyle{definition}
\newtheorem{remark}[theorem]{Remark}

\newtheorem{assumption}{Assumption}[section]

\newcommand{\smfrac}[2]{{\textstyle \frac{#1}{#2}}}
\def\<{\langle}
\def\>{\rangle}
\def\Etfw{E^{\rm TFW}}

\newcommand{\R}{\mathbb{R}}  
\newcommand{\Z}{\mathbb{Z}}  
\newcommand{\N}{\mathbb{N}}  

\newcommand{\LL}{\mathbb{L}}

\newcommand{\bfk}{{\bf k}}

\newcommand{\bfn}{{\bf n}}

\newcommand{\bfr}{{\bf r}}

\newcommand{\bfx}{{\bf x}}
\newcommand{\bfy}{{\bf y}}
\newcommand{\bfz}{{\bf z}}
\newcommand{\bfX}{{\bf X}}
\newcommand{\bfY}{{\bf Y}}

\newcommand{\bfR}{{\bf R}}

\newcommand{\ri}{{\mathrm i}}

\newcommand{\cE}{{\mathcal E}}

\newcommand{\cJ}{{\mathcal J}}
\newcommand{\cP}{{\mathcal P}}

\newcommand{\cS}{{\mathcal S}}

\newcommand{\1}{{\mathds 1}}

\newcommand{\veps}{\varepsilon}

\newcommand\per{ {\rm per}}
\newcommand\loc{ {\rm loc}}

\newcommand\Tr{{\rm Tr} \,}
\newcommand\VTr{\underline{\rm Tr} \,}
\newcommand\TrL{{\rm Tr}_L \,}
\newcommand\KS{{\rm KS}}
\newcommand\HF{{\rm HF}}
\newcommand\rHF{{\rm rHF}}

\newcommand\tot{{\rm tot}}
\newcommand\cR{\mathscr{R}}

\newcommand\ma{m_{\rm a}}
\newcommand\mnuc{m^{\rm nuc}}
\newcommand\mnuceps{m^{\mathrm{nuc},\veps}}
\newcommand\mnucN{m^{\rm nuc}_N}

\def\ie{{\em i.e.} }
\def\eg{{\em e.g.}}

\title{Thermodynamic Limits of Electronic Systems}

\author{
   David Gontier\footnote{{\tt gontier@ceremade.dauphine.fr}; CEREMADE, Université Paris-Dauphine, PSL University, Paris, France.},
   Jianfeng Lu\footnote{{\tt jianfeng@math.duke.edu}; Departments of Mathematics, Physics, and Chemistry, Duke University, Durham, NC 27708, USA. JL was supported in part by National Science Foundation under grant NSF-DMS1454939 and NSF-CHE2037263.}
   and
   Christoph Ortner\footnote{{\tt christoph.ortner@ubc.ca};  Department of Mathematics, University of British Columbia, 1984 Mathematics Road, Vancouver, BC, Canada V6T 1Z2. CO was supported by
   ERC Starting Grant 335120 and EPSRC Grant EP/R043612/1.}
}

\date \today

\begin{document}

\maketitle

\begin{abstract}
  We review thermodynamic limits and scaling limits of electronic structure models for condensed matter. We discuss several mathematical ways to implement these limits in three models of increasing chemical complexity and mathematical difficulty: (1) Thomas-Fermi like models; (2) Hartree-Fock like models; and (3) Kohn-Sham density functional theory models.
\end{abstract}

\section{Introduction}

The goal of thermodynamic limits, as introduced in the 1960's \cite{ruelle1999statistical, georgii2011gibbs}, is to obtain mathematical models for infinite systems of particles. The overarching strategy is to study systems with a finite number of particles (which can be described efficiently by well-posed mathematical models), to let the number of particles go to infinity while filling the space, and to pass to the limit in the governing equations in order to obtain a limit model. The purpose of the present chapter is to review results of this kind in the context of electronic structure models in condensed matter.

Two prototypical applications of thermodynamic limits are (1) to justify models of the {\em energy per unit cell} of a homogeneous crystal (infinite periodic system); (2) to obtain models for the formation energy of a crystalline defects without artefacts due to the boundary conditions. In this chapter, we review different models and mathematical methods to treat both of these scenarios. Extensive references will be provided throughout the chapter.

In both cases, one would like to describe an infinite system of electrons in a potential generated by an infinite collection of nuclei at positions $\cR \subset \R^3$. In most studies, $\cR$ is a periodic lattice (we write $\cR = \LL$ in this case), or a perturbation of it, describing for instance a crystal with a defect, or a deformed crystal (there are some studies for amorphous solids, in which case $\cR$ is a random set~\cite{LahbabiDisordered}). In order to highlight the main ideas of thermodynamic limit, we restrict ourselves to the simple case of a periodic crystal with one nucleus of charge $1$ per unit cell. We denote by $\ma$ the charge density of a single nucleus, which we take to be smooth to avoid some technical details: $\ma \in C^\infty(\R^3)$ with compact support and $\int \ma = 1$. The total nuclear density of the crystal is then given by 
\begin{equation} \label{eq:def:munuc}
    \mnuc(\bfr) := \sum_{\bfR \in \cR} \ma(\bfr - \bfR).
\end{equation}
In order to approximate this infinite distribution of charges, we consider a sequence of finite systems that {\em converges} to the infinite one: we choose $\cR_N \subset \cR$ a finite subset of size $| \cR_N | = N$, and study the finite electronic problem with $N$ electrons, in the external potential generated by $N$ nuclei arranged along $\cR_N$. The total nuclear density is
\begin{equation} \label{eq:def:mnuc}
    \mnucN(\bfr) := \sum_{\bfR \in \cR_N} \ma(\bfr - \bfR).
\end{equation}

Given this finite distribution of charges, $\mnucN$, one formulates a variational problem to equilibrate the electrons,
\[
    I(\cR_N) := \inf_{\gamma_N} E(\mnucN; \gamma_N),
\]
where the infimum is taken over all admissible states $\gamma_N$ representing a system of $N$ electrons, and $E$ describes the energy of finite electronic systems. Usually, $\gamma_N$ represents the electron density or density matrix. One then aims to make various statements about the limits of the energy, $I(\cR_N)$ and the optimal electron variable $\gamma_N^0$. Examples of important questions in this context include:

\begin{itemize}
    \item {\em Does the sequence $N^{-1} I(\cR_N)$ converge to some limit $W(\cR)$ as $N \to \infty$ and $\cR_N \to \cR$?} In this case, $W(\cR)$ would correspond to an average energy per electron or energy per unit volume.
    \item {\em Does the sequence of minimisers $\gamma_N^0$ have a limit $\gamma^0$ as $N \to \infty$?} The limiting object would describe an infinite sea of electrons in a crystal. Which equations are satisfied by the limiting object $\gamma^0$?
    \item If $\cR \approx \cR'$, then {\em does the energy difference $\delta I_N := I(\cR_N) - I(\cR_N')$ have a limit $\delta I$?} If $\cR$ describes a crystal and $\cR'$ the same crystal with a local defect, then $\delta I$ is the defect formation energy.
\end{itemize}

In the following sections, we focus on the case where the energy $E$ is given by one of the following three models: the Thomas-Fermi-von Weizs\"{a}cker model in Section~\ref{sec:tfw}, the (reduced) Hartree-Fock model in Section~\ref{sec:rHF}, as well as Kohn-Sham density functional theory models, in Section~\ref{sec:dft}.

In what follows, the energy of an $N$-electron state $\gamma_N$ is denoted by $E(\gamma_N)$. The infimum of this energy if $I_N = \inf_{\gamma_N} E(\gamma_N)$, and represents the ground state energy of an $N$-electron system. The energy per unit electron is $W_N = N^{-1} I_N$. For the orbital-free TFW model the $N$-electron state is given by the electron density $\rho_N$.

\begin{remark}
    Throughout this review we are technically misapplying the term ``thermodynamic limit'', but we do so in a way that is consistent with its usage in the analysis literature. 
    In a strict sense, the thermodynamic limit was introduced to describe many-particle systems in a limit where boundary effects can be neglected, and to employ the law of large numbers, large deviation theory and ergodic theory as a transition from microscopic states to macroscopic states (variational principles, PDEs etc). 
    A key goal of this framework was to model the situation when the corresponding thermodynamic functions (pressure, free energy, susceptibility, magnetisation, etc) can have singularities which appear at the critical value of the intensive parameter (temperature, chemical potential, etc). 
    We refer to \cite{ruelle1999statistical,georgii2011gibbs} for detailed treatments of the subject.
    The connection between the present review and the classical usage of the term ``thermodynamic limit'' is the study of the {\em many-particle limit} in which boundary and domain size effects can be ignored, however there is no (genuine) statistical mechanics aspect.
\end{remark}


\section{The Thomas--Fermi--von Weizs\"{a}cker Model}
\label{sec:tfw}
Thomas--Fermi models describe electronic structure purely in terms of the
electron density and electrostatic potential, and can therefore be interpreted
as a system of two nonlinear PDEs. In this setting there is a mature theory and
general results on the structure of the model and in particular the
thermodynamic limit. The original Thomas--Fermi model, while attractive due to
its simplicity, does not allow for the existence of molecules
\cite{LiebSimon1977}. We will therefore focus on the
Thomas--Fermi--von Weizs\"{a}cker (TFW) model \cite{Von_Weizsacker1935-nj}. Our
presentation is primarily based on the monograph \cite{Catto1998_book},
but incorporates also more recent results~\cite{2015-tfw,Blanc2006-ej}.

\subsection{TFW Model for a Cluster}
\label{sec:tfw:motivation}
We consider $N$ nuclei at locations $\cR_N$ and with total charge $\mnucN$, see Eq.~\eqref{eq:def:mnuc}. The non-dimensionalised TFW energy,
parametrised by $\cR_N$ as a functional of the electron density $\rho$, is
given by
\begin{equation} \label{eq:tfw:defnE}
   \Etfw(\cR_N, \rho)
   :=
   \underset{ \text{kinetic energy}} { \underbrace{
      \int_{\R^3} \left( c_{\rm W} |\nabla \sqrt\rho |^2 + c_{\rm TF}\, \rho^{5/3} \right)
   }} \
   \underset{\text{Coulomb interaction}}{\underbrace{
      + \smfrac12 D(\rho - \mnucN, \rho - \mnucN),
   }}
\end{equation}
where we defined the Coulomb quadratic form
\begin{equation} \label{eq:coulomb kernel}
  D(f, g) :=  \iint_{(\R^3)^2} \frac{f(\bfr) g(\bfr')}{|\bfr-\bfr'|} \, d \bfr \, d \bfr'.
\end{equation}
The first two terms of~\eqref{eq:tfw:defnE} represent the kinetic energy, and the third term is the Coulomb energy. This term can further be split into
\[
    \smfrac12 D(\rho - \mnucN, \rho - \mnucN) = \smfrac12 D(\rho, \rho) - D(\rho, \mnucN) + \smfrac12 D( \mnucN,\mnucN).
\]
The first term is the direct term, or Hartree term, and descibes the mean-field self-interaction of the electrons. The second is the electron-nuclei Coulomb interaction and the last term is the nuclei-nuclei one. Since we fixed the lattice $\cR$ beforehand, the last term is constant, and does not play role in the minimisation problem.
In addition, $c_{\rm W}$ and $c_{\rm TF}$ are positive physical constants that are
irrelevant from a mathematical perspective; hence, for the sake of notational
convenience, we set them to $c_{\rm W} =
c_{\rm TF} = 1$. 

The charge-neutral electronic ground state is
obtained by solving
\begin{equation} \label{eq:tfw:ground state}
   I^{\rm TFW}(\cR_N) := \inf \big\{ \Etfw(\cR_N,\rho), \quad  \rho \geq 0, \,
               {\textstyle \int_{\R^3} \rho} = N, \, \sqrt{\rho} \in H^1(\R^3) \big\}.
\end{equation}

A direct computation shows that $\rho \mapsto \int_{\R^3}
|\nabla\sqrt\rho|^2$ is convex, which is a key ingredient to obtain the
following result (see \cite{Benguria1981} for the proof).

\begin{proposition} \label{th:tfw:benguria}
   There exists a unique minimiser $\rho_N$ of \eqref{eq:tfw:ground
   state}. In addition, $\rho_N > 0$.
\end{proposition}

It can then be readily checked, at least formally, that the minimiser satisfies the Euler--Lagrange equation
\[
   - \frac{\Delta \sqrt\rho_N}{\sqrt{\rho_N}} + \smfrac53 \rho_N^{2/3} - (\mnucN - \rho_N) \ast \frac{1}{|.|} = \theta_N,
\]
for some Lagrange multiplier $\theta_N \in \R$ associated with the charge
neutrality constraint $\int_{\R^3} \rho = N$.  It now becomes convenient to make
the transformation $\rho_N = u_N^2$, where we may again assume that $u_N > 0$,
and to introduce the total electrostatic potential
\[
   V_N^\tot := (\mnucN - \rho_N) \ast \frac{1}{|.|} - \theta_N
\]
to obtain the Euler--Lagrange system
\begin{subequations} \label{eq:tfw:EL}  \begin{align}
   \label{eq:tfw:ELu}
   &- \Delta u_N + \smfrac53 u_N^{7/3} - V_N^\tot u_N = 0, \\
   \label{eq:tfw:ELphi}
   &- \Delta V_N^\tot = 4 \pi \big( \mnucN - u_N^2 \big).
\end{align} \end{subequations}
We have absorbed the Lagrange multiplier $\theta_N$ into the electrostatic
potential $V_N^\tot$, shifting it by a constant, which in particular implies that
we need not have $V_N^\tot(\bfr) \to 0$ as $|\bfr| \to \infty$. See
\cite{Benguria1981,Catto1998_book,Lieb1997-mo} for the details of this argument.

In the remainder of our treatment of the TFW model we review results
establishing the convergence of the electron ground state $(u_N, V_N^\tot)$ as $N
\to \infty$, as the nuclei configuration $\cR_N$ grows. To establish this limit, a convenient function space setting is
provided by the spaces (we denote by $B_R(\bfr) := \{ \bfr' \in \R^3, \ | \bfr' - \bfr | < R \}$)
\[
    H^k_{\rm unif} := \big\{ v \in H^k_{\rm loc}(\R^3) \,\big|\,
            {\textstyle \sup_{\bfr \in \R^d}} \| v \|_{H^k(B_1(\bfr))} < \infty \big\}.
\]

\subsection{Thermodynamic Limit Model}
To pass to the thermodynamic limit $N \to \infty$ we begin with an {\em
infinite} collection of (smeared) nuclei at positions $\cR \subset \R^3$. Here, $\cR$ needs not be a periodic lattice. Since the energy of
an infinite system is not well-defined, the associated electronic ground state
cannot be immediately characterised by an analogue of the  variational problem
\eqref{eq:tfw:ground state}. However, the nonlinear PDE representation
\eqref{eq:tfw:EL} has a straightforward generalisation. Indeed, let $\mnuc(\bfr) :=
\sum_{\bfR \in \cR} \ma(\bfr - \bfR)$, then it is natural to suppose that the electronic ground
state for the nuclei arrangement $\cR$ is given by $\rho = u^2$, where $(u, V^\tot)$
solves
\begin{subequations} \label{eq:tfw:EL-inf}  \begin{align}
   \label{eq:tfw:ELu-inf}
   &- \Delta u + \smfrac53 u^{7/3} - V^\tot u = 0, \\
   \label{eq:tfw:ELphi-inf}
   &- \Delta V^\tot = 4 \pi \big( \mnuc - u^2 \big).
\end{align} \end{subequations}

To justify this model we will establish its well-posedness and show that
it indeed arises as the thermodynamic limit of \eqref{eq:tfw:ground state} (or,
equivalently, \eqref{eq:tfw:EL}).

To that end, we need to impose restrictions on the configuration $\cR$. We assume
that $\cR$ describes roughly uniformly distributed matter, and in particular
contains no clusters with arbitrary high densities, and no holes of arbitrary large volume. Precisely, we require that there exist constants
$c_{1,2}, C_{1,2} > 0$ such that 
\begin{align} 
    \forall \, \bfr \in \R^3, \ R > 0, \qquad 
   c_1 R^3 - c_2 \leq \# \big( \cR \cap B_R(\bfr) \big) \leq C_1 R^3 + C_2. \label{eq:tfw:nocoll_noholes}
\end{align}
This condition is equivalent to (H1) and (H2) in \cite{Catto1998_book}.
One of the main results of \cite{Catto1998_book} is the well-posedness of
\eqref{eq:tfw:EL-inf}.

\begin{theorem}[Well-Posedness {\cite[Thm 6.10]{Catto1998_book}}]
   \label{th:tfw:tdl}
   Under the condition~\eqref{eq:tfw:nocoll_noholes}, there exists a unique pair
   $(u, V^\tot) \in H^4_{\rm unif} \times H^2_{\rm unif}$, with $u \geq 0$, solving \eqref{eq:tfw:EL-inf}.
   Moreover, $\inf u > 0$.
\end{theorem}

The majority of the monograph \cite{Catto1998_book} is devoted to the proof of
Theorem~\ref{th:tfw:tdl}. Let us recall some key ideas: A crucial observation is that the linear operator
\[
   L_N \varphi := - \Delta \varphi + \big(\smfrac53 u_N^{4/3}-V^\tot_N\big) \varphi,
\]
which is a kind of linearisation of \eqref{eq:tfw:ELu}, is non-negative. This
already hints at the existence of a strong stability property. Indeed, adapting
this observation (see \textit{e.g.} the proof of \cite[Lemma 5.3]{Catto1998_book}), the
following result is shown in \cite[Thm. 3.1]{2015-tfw}, closely following
variants of the same result in \cite[Sec. 5.3]{Catto1998_book} and
\cite{Blanc2006-ej}.

\begin{lemma}[Stability and Uniqueness] \cite[Thm. 3.1]{2015-tfw}
   \label{th:tfw:stab}
   Let $\cR, \cR_* \subset \R^3$ satisfy~\eqref{eq:tfw:nocoll_noholes}, let $\mnuc_1,
   \mnuc_2$ be associated nuclear charge densities, and suppose that $(u, V^\tot),
   (u_*, V^\tot_*) \in H^4_{\rm unif} \times H^2_{\rm unif}$ are corresponding
   solutions to \eqref{eq:tfw:EL-inf} with $\inf u_i > 0$.
   Then, there exist constants $C \ge 0$ and $\alpha > 0$, depending only on $\ma$ and on the
   constants in \eqref{eq:tfw:nocoll_noholes} such that,
   \begin{equation} \label{eq:locality_estimate}
      |u( \bfr) - u_*( \bfr)| + |V^\tot( \bfr) - V^\tot_*( \bfr)|
      \leq
      C \bigg(\int_{\R^3} {\rm e}^{-\alpha | \bfr - \bfz|}  \big| \mnuc - {\mnuc_*} \big|^2(\bfz)  \, d \bfz \bigg)^{1/2}.
   \end{equation}
\end{lemma}

Lemma \ref{th:tfw:stab} immediately implies uniqueness of solutions to
\eqref{eq:tfw:EL-inf}, but it is much stronger in that it also provides a
pointwise stability that quantifies the dependence of the local electronic
structure on the far-field. We will return to this result in
\S~\ref{sec:tfw:discussion}.

To establish existence of solutions, we use a thermodynamic limit argument.
At the same time, this also justifies the model \eqref{eq:tfw:EL-inf}. To that
end, we specify a sequence of clusters approximating $\cR$: let $\cR_N \subset \cR$
and $r_N \uparrow \infty, c > 0$ such that
\begin{equation} \label{eq:tfw:condition YN}
   B_{r_N}({\bf 0}) \cap \cR \subset \cR_N \subset B_{r_N+c}({\bf 0}) \cap \cR.
\end{equation}
For each $N$, Proposition~\ref{th:tfw:benguria} yields existence and uniqueness
of an electronic ground state $(u_N, V^\tot_N)$ solving \eqref{eq:tfw:EL}.

The stability result stated in Lemma~\ref{th:tfw:stab} already hints at the
possibility of uniform {\it a priori} estimates on the solutions $(u_N,
\phi_N)$, and indeed one can prove that
\begin{equation} \label{eq:tfw:apriori estimate}
   \| u_N \|_{H^4_{\rm unif}} +
   \| V^\tot_N \|_{H^2_{\rm unif}}
   \leq C,
\end{equation}
where $C$ depends only on $\ma$ and on the constants in
\eqref{eq:tfw:nocoll_noholes}, see \cite[Prop. 3.5]{Catto1998_book} for the
(involved and technical) details. A key technical step estimating
the Lagrange multiplier, which we have hidden, is due to \cite{Solovej1990-aa}.
A summary of the proof, providing also quantitative estimates can be found in
\cite[Prop. 6.1]{2015-tfw}.

With the {\it a priori} estimate \eqref{eq:tfw:apriori estimate} in hand, we
may extract a  subsequence $(u_{N_j}, V^\tot_{N_j}) \rightharpoonup (u, V^\tot)$
weakly in $H^2_{\rm loc}$ (we say $f_j \rightharpoonup f$ weakly in $H^k_{\rm loc}$ if $f_j|_D \rightharpoonup f|_D$ weakly in $H^k(D)$ for all bounded domains $D$) and it is straightforward to deduce that the limit
satisfies the PDE \eqref{eq:tfw:EL-inf}. Since the limit is unique, it follows
that the entire sequence converges. We obtain the following
result.

\begin{theorem}[Convergence] \label{th:tfw:Ntoinf}
   Let $\cR_N, \cR$ satisfy \eqref{eq:tfw:nocoll_noholes} and
   \eqref{eq:tfw:condition YN}, and let $(u_N, V^\tot_N) \in H^4_{\rm unif} \times H^2_{\rm loc}$ and $(u, V^\tot) \in H^4_{\rm loc} \times H^2_{\rm loc}$ be the corresponding solutions of~\eqref{eq:tfw:EL} and \eqref{eq:tfw:EL-inf} respectively. Then
   \[
      (u_N, V^\tot_N) \rightharpoonup (u, V)
      \qquad \text{weakly in } H^4_{\rm loc} \times H^2_{\rm loc}.
   \]
   In particular, the convergence is locally uniform.
\end{theorem}

\subsection{Discussion}
\label{sec:tfw:discussion}
We conclude this section with a series of further remarks about possible extensions and
consequences of the results.

{\it (1) Convergence rates: } In Lemma~\ref{th:tfw:stab} the conditions on the
second solution $u_*$ can be weakened. This allows to prove that ``well
inside'' the approximate domain $\cR_N$, the solutions $(u_N, V^\tot_N)$ and $(u, V^\tot)$ are
exponentially close. Specifically, in \cite[Proposition 4.1]{2015-tfw} it is
shown that there are constants $C > 0$ and $\alpha > 0$ independent of $N \in \N$ and $\bfr \in B_{r_N}$ so that
\[
    \forall N \in \N, \ \forall \bfr \in B_{r_N}, \quad    \big|u_N(\bfr) - u(\bfr) \big| +
    \big|V^\tot_N(\bfr) - V^\tot(\bfr) \big| \leq C {\rm e}^{- \alpha \, {\rm dist}(\bfr, \partial B_{r_N}(0))}.
\]
Choosing $r_N' \ll r_N$ this readily translates into the convergence rate
\[
   \| u_N - u \|_{L^\infty(B_{r_N'})} +
   \| V^\tot_N - V^\tot \|_{L^\infty(B_{r_N'})}
   \leq C {\rm e}^{ - \alpha (r_N - r_N')}.
\]
The same argument also shows that boundary effects decay exponentially into
the bulk of the cluster and justifies the common usage of buffer regions in electronic
structure calculations.

{\it (2) Surfaces and sheets: } Our assumption \eqref{eq:tfw:nocoll_noholes}
expressly disallows configurations with large sections of vacuum, in particular
surfaces and 2D materials. Indeed, not only the mathematics but also
the underlying physics changes in such situations. We refer to
\cite{Blanc2006-ej,Lu2015-rs, CancesCao2020} for related results that go beyond
this limitation.

{\it (3) The Dirac correction: } The Thomas--Fermi--Dirac--von Weizs\"{a}cker
model adds an additional correction term to the energy functional,
\[
   E^{\rm TFDW}(\cR_N, \rho) =
    \int_{\R^3} \left( |\nabla\sqrt\rho|^2 + \rho^{5/3} \right)
   + \smfrac12 D(\rho - \mnuc_N,\rho - \mnuc_N) 
   \underset{\text{Dirac exchange}}{\underbrace{
         - c \int_{\R^3} \rho^{4/3}
   }},
\]
where the additional term can be interpreted as a model for the exchange of
energy of the electrons. The additional challenge is that $E^{\rm TFDW}$ is
no longer convex. We are unaware of an in-depth treatment of this model,
but refer to \cite[Sec 3.6.3]{Catto1998_book} for a discussion of possible
avenues and \cite{E2012-ay} for results on the related Cauchy--Born
scaling limit for this model.

{\it (5) Further orbital-free DFT models: }  Most orbital-free DFT models used in practical materials computations have more complicated functionals form for kinetic energy and exchange-correlation energy, see e.g., \cite{WangTeter:1992, WangGovindCarter:1999}. It is shown in \cite{BlancCances:2005} that the Wang-Teter kinetic energy \cite{WangTeter:1992} is not bounded from below, and thus the thermodynamic limit is ill-posed. The more complicated density-dependent orbital free kinetic-energy functionals, such as the Wang-Govind-Carter functional \cite{WangGovindCarter:1999}, are yet to be mathematically understood.

{\it (6) Charge screening: } The stability result Lemma~\ref{th:tfw:stab}
clearly shows that interaction in the TFW model is exponentially localised, {\em
despite} the presence of the long-range Coulomb interaction. This can be
interpreted as a very general screening result. Consider two configurations $\cR, \cR_*$ satisfying
\eqref{eq:tfw:nocoll_noholes}, and which coincide outside a large ball of radius $r > 0$: $\cR \setminus B_r =
\cR_* \setminus B_r$. Let $(u, V^\tot), (u_*, V^\tot_*)$ be the corresponding solutions. Then Lemma~\ref{th:tfw:stab} implies
\begin{equation} \label{eq:tfw:decay-perturbation}
   |u(\bfr) - u_*(\bfr)| + |V^\tot(\bfr) - V^\tot_*(\bfr)| \leq C {\rm e}^{- \alpha |\bfr|}.
\end{equation}
For instance, if $\cR_*$ contains more atoms than $\cR$, one could expect that the potentials satisfy $V_*^\tot(\bfr) \approx V^\tot(\bfr) + Q/| \bfr |$ as $\bfr \to \infty$, where $Q$ would be the extra effective charge. However, according to~\eqref{eq:tfw:decay-perturbation}, this is not the case: the extra charge is completely screened. This is a very general fact in TFW theory, see~\cite{Cances2011-ot, 2015-tfw} for details.

One can also take into account the relaxation of the configuration $\cR_*$ due to the presence of the defect. In this case, instead of an exponential decay, we obtain (see~\cite{ChenNazarOrtner2019})
\[
    \big|u(\bfr) - u_*(\bfr)\big| + \big|V^\tot(\bfr) - V^\tot_*(\bfr)\big|
    \lesssim |\bfr|^{-2},
\]
and, since $| \bfr |^{-2} = o(| \bfr |^{-1})$, we deduce that charges are screened, see \cite{Cances2011-ot} and \cite[Thm. 4.1]{2015-tfw} for the details.

\subsection{Scaling limit}
\label{sec:tfw:scaling limit}
A question related to but distinct from the thermodynamic limit arises when
considering the derivation of a continuum model for elastic material response
from an underlying electronic structure model. Such scaling limits for the TFW
model were first studied in \cite{BlancLeBrisLions:2002}, but the following
discussion is builds on the results of
\cite{2015-tfw}. Specifically, the stability and locality estimate
\eqref{th:tfw:stab}  yields stronger and quantitative results. In
addition, for the sake of consistency with the KS-DFT case in Section~\ref{sec:dft}, we consider a
periodic instead of an infinite-domain setting.

\paragraph{Spatial decomposition of energy}
In preparation we first mention another useful consequence of the stability and
locality estimate found in Lemma~\ref{th:tfw:stab}, which is also of independent interest:
Let $\cR \subset \R^3$ be a finite configuration of nuclei or an infinite
configuration satisfying \eqref{eq:tfw:nocoll_noholes} and let $(u, V^\tot)$ be
the associated solutions to \eqref{eq:tfw:EL}, then we define the energy density
\[
    \mathcal{E}(\cR;\cdot)
        = |\nabla u|^2 + u^{10/3} + \frac{1}{8\pi} |\nabla V^\tot|^2.
\]
If $\cR$ is finite then one may readily check \cite[Eq. 4.18]{2015-tfw}
 that $I^{\rm TFW}(\cR) = \int_{\R^3} \mathcal{E}(\cR; \bfr) \,d\bfr.$
We may therefore think of
\[
    I^{\rm TFW}(\cR, \Omega) := \int_\Omega \mathcal{E}(\cR;\bfr) \, d\bfr
\]
as the energy stored in a compact sub-domain $\Omega \subset \R^3$. This
intuition is further supported by the following result, proven in
\cite[Proof of Thm. 4.2]{2015-tfw}, which is closely related
to \eqref{th:tfw:stab}: there exist constants $C, \gamma$ such that
\begin{equation} \label{eq:tfw:loc_Edens}
   \forall \bfR \in \cR, \ \forall \bfr \in \R^3, \qquad  \left|\frac{\partial \mathcal{E}(\cR; \bfr)}{\partial \bfR} \right|
    \leq C e^{-\alpha| \bfr - \bfR|}.
\end{equation}
In \cite[Sec. 4.4]{2015-tfw} this observation is used to demonstrate exponential
locality of interatomic forces in the TFW model. In the following section we use it for an easy derivation of the Cauchy--Born scaling limit.

\paragraph{The Cauchy--Born scaling limit}
Consider a periodic arrangement $\cR = B \Z^3$ of nuclei, where $B \in \R^{3 \times 3}$ is a non-singular matrix, and let $(u,V^\tot)$ describes the corresponding TFW ground-state.
Then, uniqueness of solutions to \eqref{eq:tfw:EL} implies that they must observe
the same periodicity. In particular, we can define the Cauchy--Born energy function, which represents the energy stored in $\Omega_B := B [0, 1)^3$, that is
\[
W^{\rm cb}(B) :=  I^{\rm TFW}(B \Z^3, \Omega_B) := \int_{\Omega_B} \mathcal{E}(\cR; \bfr) \,d\bfr.
\]
A deformed configuration of the crystal is described by a continuum deformation field $Y(\bfx) = \bfx + U(\bfx)$  where $U$ is smooth and $\cR$-periodic. We assume that $U$ is chosen such that $Y$ is bijective, \textit{i.e.} a proper deformation.

\begin{figure}
    \centering
    \includegraphics[width=\textwidth]{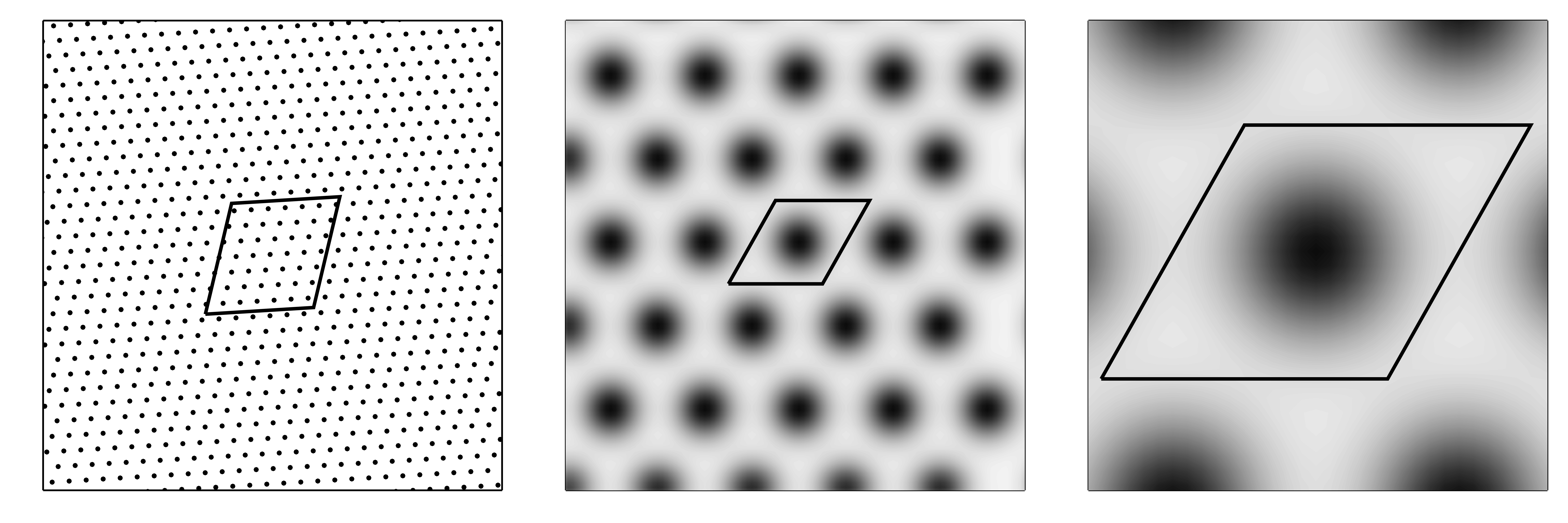}
    \caption{Illustration of the scaling limit. Left: a deformation of a homogeneous crystal varying slowly relative to the scale of atoms; Center: the blow-up of a small section is nearly homogeneous; Right: The near-homogeneous section can be approximately represented by a single unit cell.}
    \label{fig:scalinglimit}
\end{figure}

Given a parameter $\epsilon > 0$ describing the inverse length-scale over which
the deformation varies we define a deformed crystalline configuration by
\[
    \cR^\epsilon := \big\{ Y_\epsilon(\bfx) := \epsilon^{-1} Y\big(\epsilon \bfx\big) \,|\,
            \bfx \in B \mathbb{Z}^3 \big\}.
\]
This definition encodes the {\em Cauchy--Born hypothesis} that nuclei in a
crystal follow the continuum deformation field. An example of such an atomistic configuration shadowing a continuum field is given in Figure~\ref{fig:scalinglimit}(left). It must be emphasised that this
is a simplifying assumption that is only approximately valid in specific
deformation regimes and for simple crystals; see
\cite{Weinan2007-zd,Ortner2013-en} for in-depth discussions.

In the following we concern ourselves with the {\em scaling limit} $\epsilon \to
0$ of the stored elastic energy per unit undeformed volume. To that end, let $\Omega_\epsilon :=
\epsilon^{-1} \Omega$ then $\cR^\epsilon$ has periodic cells $\Omega_\epsilon$
or alternatively also $Y_\epsilon(\Omega_\epsilon)$. That is, we may write
the elastic energy per unit volume as
\[
    E^\epsilon :=
    |\Omega^\epsilon|^{-1} I^{\rm TFW}(\cR^\epsilon, \Omega_\epsilon)
    =
    \frac{1}{|\Omega^\epsilon|} \int_{\Omega^\epsilon}
        \mathcal{E}(\cR^\epsilon, \bfr) \, d\bfr
    =
    \frac{\epsilon^3}{\det B} \int_{Y_\epsilon(\Omega_\epsilon)}
        \mathcal{E}(\cR^\epsilon, \bfr) \, d\bfr.
\]
We remark that the electronic coordinates $\bfr$ belong to the deformed space
\textit{i.e.} it is natural to write $\bfr = Y_\epsilon(\bfx)$.

The key observation now is that the locality estimate \eqref{eq:locality_estimate} on the
electronic structure and its extension to the energy density
\eqref{eq:tfw:loc_Edens} suggest that to predict the value of
$\mathcal{E}(\cR^\epsilon, \bfr)$ for a non-uniform deformation varying at the macroscopic scale, it is not required to be aware of the global
configuration $\cR^\epsilon$ but it is sufficient to know the {\em local
deformation} near $\bfr$. This is illustrated by Figure~\ref{fig:scalinglimit}.

To make this precise, let $\bfn \in \mathbb{L}$,
$t_{\bfn} = Y_\epsilon(\bfn)$ and $F_{\bfn} =
\nabla Y_\epsilon(\bfn)$, then
\[
    Y_\epsilon(\bfx) = t_{\bfn} + F_{\bfn} (\bfx - \bfn)
     + O(\epsilon),
\]
for $\bfx$ in any bounded neighbourhood of $\bfn$. A Taylor expansion of
$\mathcal{E}$ with respect to $\cR^\epsilon$, employing
\eqref{eq:tfw:loc_Edens}, implies that for $\bfr \in Y_\epsilon(\bfn+\Omega)$ we
have
\[
    \mathcal{E}(\cR^\epsilon, \bfr) =
    \mathcal{E}\big( F \Z^d, \bfr \big) + O(\epsilon).
\]
After integrating over one cell in the undeformed crystal, which in deformed
coordinates becomes $Y_\epsilon(\bfn + \Omega)$, and making further elementary
approximation, we obtain
\[
    \int_{Y_\epsilon(\bfn + \Omega)} \mathcal{E}(\cR^\epsilon, \bfr)\,d\bfr
    = W(F_\bfn) + O(\epsilon).
\]

Finally, after summing over all such unit cells, using the scale-invariance of
the deformation gradient one obtains the following convergence result. The
second order error $O(\epsilon^2)$ is obtained by a more careful exploitation of
the point symmetry in simple crystal lattices.

\begin{theorem}
  Let $Y \in C^4_{\rm per}(\Omega)$, then
  \[
      E^\epsilon(Y) = \int_\Omega W^{\rm cb}(\nabla Y(\bfx)) \, d\bfx
          + O(\epsilon^2)
      \qquad \text{as } \epsilon \to 0.
  \]
\end{theorem}
The presentation of this section follows unpublished notes. Related results
using different techniques were first presented in \cite[Theorem 5, case
(i)]{BlancLeBrisLions:2002}. The publication \cite{BlancLeBrisLions:2002} also
consider several generalisations, including domains with boundaries and
alternative scaling regimes.

\section{The reduced Hartree-Fock model}
\label{sec:rHF}

We now focus on the reduced Hartree-Fock (rHF) model. In this model, a fermionic system with $N$ electrons is described by a one-body density matrix $\gamma$, which is a self-adjoint operator on $L^2(\R^3)$ satisfying the Pauli principle $0 \le \gamma \le 1$, and with trace $N$.

Together with the spectral theorem, this implies that $\gamma$ is of the form
\[
    \gamma = \sum_{i=1}^\infty n_i | \phi_i \rangle \langle \phi_i |, \quad \text{with} \quad 1 \ge n_1 \ge n_2 \ge \cdots \ge 0,
    \quad \text{and} \quad \Tr(\gamma) = \sum_{i=1}^\infty n_i = N.
\]
Here, the functions $(\phi_i)_{i}$ form an orthonormal basis of eigenfunctions in $L^2(\R^3)$, and are called the {\em orbitals}, and the numbers $0 \le n_i \le 1$ are the {\em occupation number}. 
To such a one-body density matrix, we can associate its density $\rho_\gamma(\bfr) := \gamma(\bfr, \bfr) = \sum_{i} n_i | \phi_i |^2(\bfr)$. 

In the potential generated by the nuclei at $\cR_N \subset \R^3$, the rHF energy of a state $\gamma$ is
\begin{equation} \label{eq:def:ErHF}
E^\rHF(\cR_N, \gamma) :=  \frac12 \Tr (- \Delta \gamma) + \frac12 D(\rho_\gamma - \mnuc_N, \rho_\gamma - \mnuc_N),
\end{equation}
where $\mnuc_N$ is the total nuclear density defined in~\eqref{eq:def:munuc}. Compared with~\eqref{eq:tfw:defnE}, we see that the kinetic energy part $\int | \nabla \sqrt{ \rho } |^2 + \rho^{5/3}$ has been replaced by 
\[
    \frac12 \Tr( - \Delta \gamma) := \frac12 \sum_{i=1}^\infty n_i \| \nabla \phi_i \|_{L^2(\R^3)}^2.
\]
In particular, this model is no longer a function of the density $\rho$, but of the one-body density matrix $\gamma$. The energy of the configuration $\cR_N$ is given by the minimisation problem
\begin{equation} \label{eq:def:Energy(r)HF}
    I^{\rHF}(\cR_N) := \inf \left\{ E^\rHF(\cR_N, \gamma), \ 0 \le \gamma = \gamma^* \le 1, \ \Tr(\gamma) = N  \right\}.
\end{equation}

Closely related the to rHF model is the Hartree-Fock (HF) model, where the exchange term is considered. This term is a correction to the direct Hartree energy, and is due to the fermionic nature of the particles. The HF model reads
\begin{equation} \label{eq:def:EHF}
    E^\HF(\cR_N, \gamma) :=  E^\rHF(\cR_N, \gamma) -  \frac12 \iint_{(\R^3)^2} \dfrac{| \gamma (\bfr, \bfr') |^2}{| \bfr - \bfr' |}  d \bfr d \bfr'.
\end{equation}
Since the HF model is not convex in $\gamma$, we only have partial results for the HF model, and most of the following facts only hold for the rHF model.

In the thermodynamic limit, we consider a regular periodic lattice $\LL \subset \R^3$, and a sequence of arrangements $\cR_N \subset \LL$ with $| \cR_N | = N$, and satisfying~\eqref{eq:tfw:condition YN}. We want to study the energy per unit cell $N^{-1} I^\rHF(\cR_N)$ as $N$ goes to infinity.

\medskip
 
The finite electron model~\eqref{eq:def:Energy(r)HF} was introduced and studied by Solovej~\cite{solovej1991proof}. The existence of an optimiser $\gamma_N^0$ is provided here.
The thermodynamic limit was latter studied by Catto, Le Bris and Lions in a series of paper. In~\cite{Catto1998}, the authors announced their results, latter proved in~\cite{Catto2001} (for the models presented here) and~\cite{Catto2002} (for the pure-state version of these problems, \ie when $\gamma$ is further constrained to be a rank-$N$ projector). They prove the thermodynamic limit for the rHF model:
\begin{equation} \label{eq:thermoCatto2001}
    \lim_{N \to \infty} \frac{1}{N} I^\rHF(\cR_N)  = I^\rHF_\per + \frac{\mathfrak{m}}{2},
\end{equation}
where $I^\rHF_\per$ can be characterised by a minimisation periodic problem, that we describe in the next section, and $\mathfrak{m}$ is the Madelung constant, see~\eqref{eq:def:Madelung} below. They conjectured that a similar result should hold for the HF model. Finally, they proved that the limiting problem $I^\rHF_\per$ (and its HF counterpart $I^\HF_\per$) is indeed well-posed. We discuss this point in the next section.

\subsection{The periodic model}\label{sec:periodicHF}

In order to write the limiting periodic model, as introduced by Catto, Le Bris and Lions, we define the set of periodic one-body density matrices (recall that in our simple setting, we expect one electron per unit cell)
\begin{equation*}
    \mathcal{P}_{\per} := \left\{  0 \le \gamma = \gamma^* \le 1, \ \forall \ell \in \LL, \ \tau_\ell \gamma = \gamma \tau_\ell, \ \VTr \gamma = 1   \right\},
\end{equation*}
where $\tau_\ell$ is the translation operator $\tau_\ell f(\bfx) := f(\bfx - \ell)$. Such a periodic density matrix has a $\LL$-periodic density $\rho_\gamma(\bfr) := \gamma(\bfr, \bfr)$. Its trace per unit cell $\VTr \gamma$ is defined by
\[
    \VTr \gamma := \Tr \left( \1_\Omega \gamma  \1_\Omega \right) = \int_{\Omega} \rho_\gamma,
\]
where $\Omega$ is a unit cell associated to the lattice $\cR = \LL$. A periodic density matrix $\gamma \in \mathcal{P}_\per$ has a $\LL$-periodic density $\rho_\gamma (\bfx) := \gamma(\bfx, \bfx) \in L^1_\per(\Omega)$. We let $G$ be the $\LL$-periodic Coulomb kernel, solution to
\[
    - \Delta G := 4 \pi \sum_{\bfR \in \LL} \left( \delta_\bfR - | \Omega |^{-1} \right), \quad \text{and} \quad \int_{\Omega} G = 0,
\]
and we introduce the periodic Coulomb quadratic form $D_\per(\cdot, \cdot)$ defined for periodic functions by (compare with~\eqref{eq:coulomb kernel})
\[
    D_\per(f, g) :=  \iint_{(\Omega)^2} f(\bfr) g(\bfr') G(\bfr - \bfr') d \bfr d \bfr'.
\]

The Madelung constant appearing in~\eqref{eq:thermoCatto2001} is defined to be
\begin{equation} \label{eq:def:Madelung}
    \mathfrak{m} := \lim_{\bfr \to {\bf 0}} \left( G(\bfr) - \frac{1}{| \bfr |} \right).
\end{equation}
Note that since $F(\bfr) := G(\bfr) -  | \bfr |^{-1}$ satisfies $\Delta F = 0$ on $\Omega$, the function $F$ is indeed smooth on $\Omega$, hence has a well-defined value at $\bfr = {\bf 0}$. The Madelung constant somehow describes the mismatch between the full space Coulomb kernel $| \bfr |^{-1}$, and the periodic one $G(\bfr)$ (which can {\em a priori} be defined up to a constant).

With these notations, the limit $W_\per^\rHF$ for the perfect crystal is defined as the minimisation problem~\cite{Catto2001}
\begin{equation} \label{eq:def:IrHF_per}
    W_\per^\rHF := \inf \left\{ E^{\rHF}_\per (\LL, \gamma), \ \gamma \in \mathcal{P}_\per \right\},
\end{equation}
where the energy per unit cell $E^\rHF_\per$ is
\begin{equation} \label{eq:def:cIrHF_per}
    E^\rHF_\per(\LL, \gamma) := \frac12 \VTr(- \Delta \gamma) + \frac12 D_\per(\rho_\gamma - \mnuc_\per, \rho_\gamma - \mnuc_\per),
\end{equation}
and where $\mnuc_\per(\bfx) := \sum_{\ell \in \LL} \ma(\bfx - \ell)$ is the periodic nuclear density. Comparing this expression with~\eqref{eq:def:ErHF}, we see that all terms have been <<normalised>> to take into account the periodicity of the infinite system.

\medskip

The fact that $I^\rHF_\per$ is a well-posed problem was proved by Catto, Le Bris and Lions in~\cite{Catto2001}. Later in~\cite{Cances2008}, Cancès, Deleurence and Lewin proved that the minimiser $\gamma$ satisfied the Euler-Lagrange equations
\[
\gamma = \1 \left( H_\gamma \le \veps_F \right), \quad H_\gamma := - \frac12 \Delta + (\rho_\gamma - \mnuc_\per)*G.
\]
Here, $\veps_F \in \R$ is the {\em Fermi level}. The operator $H_\gamma$ is the mean-field one-body Hamiltonian of the crystal, which is a self-adjoint operator that commutes with $\LL$-translations. Its spectral properties are well understood thanks to the Bloch transform~\cite[Chapter XIII]{ReedSimon4}, and its spectrum are composed of bands and gaps. When $\veps_F$ is in a gap, the crystal is an insulator, while when $\veps_F$ is in a band, it is a metal.

\subsection{Supercell methods, and periodic thermodynamic limit}

Once the periodic problem~\eqref{eq:def:IrHF_per} has been written and justified, it is possible to understand its properties from other approaches. In~\cite{Cances2008} (see also~\cite{Deleurence2008_thesis}), Cancès, Deleurence and Lewin proved that this problem was also the limit of another thermodynamic limit. Their idea was to start directly with a periodic problem on the large supercell $\Omega_L :=  L \Omega$ with $N = L^3$ electrons, and take the limit $L \to \infty$. In other words, instead of working with one-body density matrices $\gamma$ acting on the whole space $L^2(\R^3)$, they looked at one-body density matrices acting on the {\em supercell} $L^2_{\rm per}(\Omega_L)$. We therefore define
\[
    \cP_\per^L := \left\{  \gamma \ \text{acting on} \  L^2_{\rm per}(\Omega_L), \ 0 \le \gamma = \gamma^* \le 1, \  \TrL \gamma = L^3  \right\},
\]
where we set for simplicity $\TrL := \Tr_{\cS(L^2_{\rm per}(\Omega_L))}$. A one-body operator $\gamma \in \cP_\per^L$ has an $L\LL$-periodic density $\rho_\gamma \in L^1_\loc$. We also define the $L \LL$-periodic Coulomb kernel as $G_L(\bfx) := L^{-1} G(L^{-1} \bfx)$, and the $L$-periodic Coulomb quadratic form defined for $L\LL$-periodic functions by
\[
    D_L(f, g) := \iint_{(\Omega_L)^2} f(\bfr) g(\bfr) G_L(\bfr - \bfr') d \bfr d \bfr'.
\]

The supercell model is given by a periodic minimisation problem of the form
\begin{equation} \label{eq:def:ErHF_supercell}
    I^\rHF_{\per,L}(\cR^L) := \inf \left\{ E^\rHF_{\per,L}(\cR^L, \gamma), \ \gamma \in \cP_\per^L \right\},
\end{equation}
with the supercell energy
\begin{equation} \label{eq:def:cErHF_supercell}
    E^\rHF_{\per,L}(\cR^L, \gamma) := \frac12 \TrL (- \Delta_L \gamma) + \frac12 D_L(\rho_\gamma - \mnuc_L, \rho_\gamma - \mnuc_L).
\end{equation}
Here, $\cR^L$ is an $L \LL$-periodic lattice (for instance $\cR^L = \LL$, or a deformation of it, see below), and $\mnuc_L$ is the nuclear density $\mnuc_L := \sum_{\bfR \in \cR_L} \ma(\bfx - \bfR)$. In the case $\cR^L = \LL$, there are $L^3$ nuclei and electrons per supercell.

Even in the perfect crystal case, that is when $\cR^L = \LL$, the problems~\eqref{eq:def:IrHF_per}-\eqref{eq:def:cIrHF_per} and~\eqref{eq:def:ErHF_supercell}-\eqref{eq:def:cErHF_supercell} differ. In~\eqref{eq:def:IrHF_per}, the minimisation is performed for $\gamma$ acting on the whole space $L^2(\R^3)$, while in~\eqref{eq:def:ErHF_supercell}, it is performed for $\gamma$ acting on the supercell $L^2(\Omega_L)$. These two types of operators cannot be compared, and it is not obvious {\em a priori} that there is a link between the two problems. Still, both operators give $\LL$-periodic densities, which can be compared. This important fact allows to prove the convergence~\cite{Cances2008}
\begin{equation} \label{eq:limEL}
    \lim_{L \to \infty} \dfrac{1}{L^3} I^\rHF_{\per,L}(\LL) = W^\rHF_{\per}.
\end{equation}
The result was later refined in~\cite{Gontier2016convergence}, where the authors proved that, in the insulating case (see~\cite{Cances2020numerical} for the metallic case), the convergence is exponential, in the sense that there exist constants $C \in \R$ and $\alpha > 0$ such that
\begin{equation} \label{eq:speedCV_EL}
    \left| W_\per^\rHF - \frac{1}{L^3} I^\rHF_{\per, L}(\LL) \right| + \| \rho^0_{\rm per} - \rho^0_{\rm per, L} \|_{L^\infty} \le C {\rm e}^{-\alpha L},
\end{equation}
where $\rho^0_\per$ and $\rho^0_{\per,L}$ are the electronic densities of the periodic and supercell minimisers respectively, seen here as $\LL$-periodic functions. This exponential\textbf{} convergence comes from the analyticity of the Bloch representation.

This means that the full space problem $W^\rHF_\per$ can be well-approximated by the supercell model $I^\rHF_{\per, L}$. This latter problem can be studied efficiently from a numerical point of view, thanks to the Bloch transform. As noticed in~\cite{Gontier2016convergence}, the supercell model corresponds exactly to a uniform discretisation of the Brillouin zone, as described in a famous paper by Monkhorst~\cite{Monkhorst1976}. For metallic systems, the exact rate of convergence is unknown in the general case (see~\cite{Cances2020numerical} for details).

\begin{remark} \label{rem:symBreaking}
    When studying supercell methods for non convex problems, symmetry breaking may happen (see \eg~\cite{Ricaud2018, GonLewNaz-21}). In this case, the density of the $L \LL$-periodic problem may not be $\LL$-periodic, and the periodic problem may not be the limit of supercell models. 
\end{remark}

To sum up, the energy per unit cell $W_\per^\rHF$ is the limit of two different sequences, namely
\[
    W_\per^\rHF = \lim_{N \to \infty} \left(  \frac{1}{N} I^\rHF(\cR_N) \right) - \frac{\mathfrak{m}}{2} ,
    \quad \text{and} \quad
    W_\per^\rHF = \lim_{L \to \infty} \left( \frac{1}{L^3} I^\rHF_{\per, L}(\LL) \right).
\]
In the first limit, the crystal is seen as the limit of finite systems. This is the correct physical limit, as a real crystal is indeed always finite. However, we expect the convergence to be slow, due to boundary effects. On the other hand, the second limit has no real physical meaning, but gives exponential rate of convergence in the insulating case. 

\subsection{Local defects in crystals, in the reduced Hartree-Fock model}

We now discuss how to define the energy of a defect inside a crystal. We would like to define this energy as the difference between the energy of a crystal with a defect, and the energy of the crystal without the defect. Unfortunately, these two quantities are infinite. Also, the model with defect does not have an underlying periodicity, hence there is no notion of {\em energy per unit cell} in this case. One way to define the energy of a defect is through a thermodynamic limit procedure.

\medskip

Let $\cR := \LL$ be the arrangement of nuclei for the perfect crystal, and let $\cR_*$ be the one for the crystal with (local) defect, that is such that $\cR$ and $\cR_*$ coincide outside a ball of radius $r > 0$. The nuclear charge of the defect is therefore
\[
    \nu := \mnuc_* - \mnuc = \sum_{\bfR \in \cR_* \setminus \cR} \ma(\cdot - \bfR) - \sum_{\bfR \in \cR \setminus \cR_*} \ma(\cdot - \bfR).
\]

For $L \in \N^*$, we can consider $\cR_*^L$ the $L \LL$-periodic arrangement which is equals to $\cR_*$ on $\Omega_L$ (note that $\cR^L = \cR = \LL$). In~\cite{Cances2008}, the authors consider the supercell energy of the defect $\nu$, defined by
\[
    J_L(\nu) := I^\rHF_{\per, L}(\cR_*^L) - I^\rHF_{\per,L}(\LL).
\]
Here there is a slight complication: since we do not know {\em a priori} how many electrons should be in the system, we should not fix the number of electrons, but rather the Fermi level (grand canonical ensemble). We do not comment on this point to keep this presentation simple, and refer to~\cite{Cances2008} for details.

Although the two quantities $I^\rHF_{\per, L}(\cR_*^L)$ and $I^\rHF_{\per,L}(\LL)$ diverges to infinity with rate $O(L^3)$, the difference of the two quantities stays finite in the limit, and we can define the energy of the defect as
\[
    J_\infty(\nu) := \lim_{L \to \infty} J_L(\nu).
\]
In~\cite{Gontier2016supercell}, the authors prove that the corresponding rate of convergence is $O(L^{-1})$. This slow rate of convergence is due to the spurious interaction between the defect and its periodic images, a fact predicted in~\cite{Leslie1985, Makov1995}. This makes the supercell method quite a poor numerical method in this case. 

It turns out that the limit $\cJ_\infty(\nu)$ can be characterised as a minimisation problem on a set of "defect" operators. It is unclear whether this last problem could be tackled directly with efficient numerical methods (see also~\cite{Cances2008non}).


\section{Scaling limit for Kohn-Sham DFT}
\label{sec:dft}

In this section, we discuss Kohn-Sham models. For a finite system with $N$ electrons described by a one-body density matrix $\gamma$, the Kohn-Sham density functional takes the form
\begin{equation}\label{eq:KS}
  \cE^{\KS}\bigl(\cR_N, \gamma \bigr)
  := \frac{1}{2}\Tr( - \Delta \gamma ) +
  \frac{1}{2}D(\rho_\gamma - \mnuc_N, \rho_\gamma - \mnuc_N) + E_{\text{xc}}[\rho_{\gamma}],
\end{equation}
where $\mnuc_N$ is the total nuclear density defined in~\eqref{eq:def:munuc}. Compared with the reduced Hartree-Fock model~\eqref{eq:def:ErHF}, the Kohn-Sham model includes the exchange-correlation energy $E_{\text{xc}}[\rho_{\gamma}]$, where we have adopted the notation for a LDA or GGA type functional
and thus it can be explicitly written in terms of $\rho_{\gamma}$ as
\begin{align*}
  & \text{(LDA)} \qquad E_{\text{xc}}[\rho_{\gamma}] = \int_{\mathbb{R}^3} \epsilon_{\text{xc}}(\rho_{\gamma}(\bfr)) d\bfr, \qquad \text{or} \\
  & \text{(GGA)} \qquad E_{\text{xc}}[\rho_{\gamma}] = \int_{\mathbb{R}^3}
    \epsilon_{\text{xc}}\bigl(\rho_{\gamma}(\bfr), \bigl\lvert \nabla \sqrt{\rho_{\gamma}(\bfr)} \bigr\rvert^2\bigr) d\bfr.
\end{align*}
Even the simplest exchange-correlation functionals used in practice
has complicated expressions, and hence will not be given explicitly
here. Most of them are non-convex in $\rho$, as, for instance the
Dirac exchange term $E_{\mathrm{x}}^{\text{Dirac}}[\rho] = - C_D\int \rho(\bfx)^{4/3} d\bfx$. Thus, even
the existence of minimiser to the Kohn-Sham DFT problem becomes a difficult
question. The existence of minimisers for LDA functionals has been
proved in \cite{LeBris:1993, AnantharamanCances:2009, Gontier:2014},
while for GGA type functionals, it remains open with only 
preliminary results available (see the case of $N = 1$ in~\cite{AnantharamanCances:2009}).

We will henceforth assume LDA type exchange-correlation functionals.
The variation of the functional \eqref{eq:KS} gives rise to the
Kohn-Sham equations for  $\gamma = \sum_{i=1}^N | \psi_i \rangle \langle \psi_i |$, where $\psi_i$ are the Kohn-Sham orbitals, solution to
\begin{equation}\label{eq:KohnShamEqn}
  H^\KS[\rho_\gamma] \psi_i = E_i \psi_i
   \quad \text{where} \quad
  H^{\KS}[\rho] :=  -\frac{1}{2} \Delta + V_{\text{H}}[\rho] + V_{\text{xc}}[\rho]
\end{equation}
with $\rho_{\gamma}$ the density associated with $\gamma$ and the Hartree and exchange-correlation potentials respectively given by
\begin{equation*}
      V_{\text{H}}[\rho] = (\rho - \mnuc_N) * \frac{1}{| \cdot |}
      \quad \text{and} \quad
      V_{\text{xc}}[\rho] := \epsilon_{\text{xc}}'(\rho(\cdot)).
\end{equation*}
The Kohn-Sham equations \eqref{eq:KohnShamEqn} is a set of nonlinear
eigenvalue problems, as the effective Hamiltonian operator
$H^{\KS}[\rho_\gamma]$ depends on the solution $\gamma$. We remark that in general there is no guarantee that
the Kohn-Sham orbitals of the minimisers of \eqref{eq:KS} correspond to the lowest $N$
eigenvalues of the self-consistent Hamiltonian, though in practice
this is often assumed and known as the \textit{Aufbau principle}.

Due to the non-convexity and hence possible symmetry breaking, see Remark~\ref{rem:symBreaking}, the thermodynamic limit of Kohn-Sham DFT with exchange-correlation functionals is very challenging and not much progress has been made.

To understand the behaviour of electronic structure in materials, we
take a typical starting point of modelling in materials science -- the periodic Kohn-Sham model with supercell $\Omega$. This can be
formulated using the density matrix similar to the periodic
Hartree-Fock model discussed in Section~\ref{sec:periodicHF}. 
A periodic Kohn-Sham energy is of the form
\[
    E^{\rm KS}_{\rm per}(\LL, \gamma) = E^{\rm rHF}_{\rm per}(\LL, \gamma) + E^{\rm xc}_{\rm per}[\rho_\gamma],
\]
where the rHF energy $E^{\rm rHF}_{\rm per}$ was defined in~\eqref{eq:def:cIrHF_per}. This is the rHF model with the addition of a periodic exchange-correlation energy. One could follow the lines of Section~\ref{sec:periodicHF} to study the thermodynamic limit. We can also consider the following alternative formulation, presented in~\cite{ELu:2013}, and that we present now.

The self-consistent Kohn-Sham eigenvalue problem
\eqref{eq:KohnShamEqn} can be reformulated as a fixed point equation
for the density
\begin{equation}\label{eq:kohnshammap}
  \rho(\bfr) = \mathcal{F}^{\KS}[
  \rho](\bfr) :=
  \Biggl[\frac{1}{2\pi \ri}\oint_{\mathscr{C}} \bigl(\lambda - H^{\KS}[\rho] \bigr)^{-1} d \lambda \Biggr] (\bfr, \bfr),
\end{equation}
where $\mathscr{C}$ is a contour in the resolvent set separating the
first $N$ eigenvalues of $H^{\KS}$ from the rest of the spectrum
(assuming a spectral gap). The right hand side of
\eqref{eq:kohnshammap} denotes the diagonal of the kernel of the
density matrix viewed as an integral operator.

\subsection{Periodic Kohn-Sham DFT model}
\begin{figure}[ht]
    \centering
    \includegraphics[width=\textwidth]{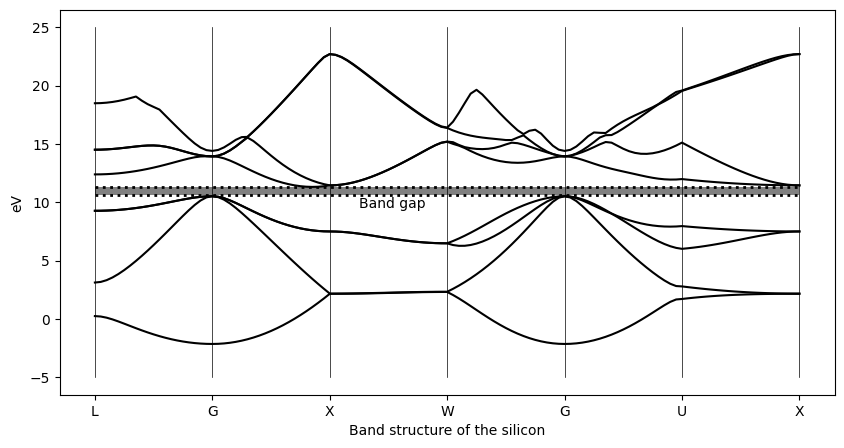}
    \caption{Schematic band structure of crystalline  silicon, along various lines connecting high-symmetry points in the first Brillouin zone $\Omega^{\ast}$. The first $4$ bands are occupied and separated by a band gap from the higher bands.}
    \label{fig:bandstructure}
\end{figure}

For a periodic system with Bravais lattice $\LL$, we can write a similar equation. We introduce the periodic Kohn-Sham Hamiltonian associated with some $\LL$-periodic density
 $\rho$, given
by
\begin{equation*}
  H^{\KS}_{\per}[\rho] = - \frac{1}{2} \Delta + V_{\text{H}, \per}[\rho] + V_{\text{xc}}[\rho],
\end{equation*}
where the periodic Hartree potential solves
\begin{equation*}
  -\Delta V_{\text{H}, \per}[\rho] = 4\pi ( \rho - \mnuc)
\end{equation*}
with periodic boundary condition where $\mnuc$ is understood as a
background charge density given by the nuclei (to be specified
below). As the potential is periodic, the Bloch-Floquet theory applies
to the Hamiltonian. In particular, the spectrum of
$H^{\KS}_{\per}[\rho]$ has a band structure. For each
$\bfk \in \Omega^{\ast}$ the first Brillouin zone, the Bloch waves
solve the eigenvalue problem
\begin{equation*}
  \Biggl(\frac{1}{2} \bigl( - \ri \nabla + \bfk \bigr)^2 + V_{\text{H}, \per}[\rho] + V_{\text{xc}}[\rho] \Biggr)
u_{n, \bfk}(\bfx)
  = E_{n, \bfk} u_{n, \bfk}(\bfx), \quad \forall n = 1,2,\ldots\textbf{},
\end{equation*}
with periodic boundary condition on $\Omega$. The spectrum is given by
\begin{equation*}
  \sigma(H^{\KS}_{\per}[\rho]) = \bigcup_{n} \bigcup_{\bfk \in \Omega^{\ast}} E_{n, \bfk}.
\end{equation*}
This is known as the band structure, see Figure~\ref{fig:bandstructure} for an illustration.

If the first $N$ bands are occupied and there exists a gap between the
occupied and unoccupied spectrum (in physical terms, the system is an
insulator), the Kohn-Sham map can be generalised to the periodic
setting as
\begin{equation}\label{eq:perKohnShammap}
  \mathcal{F}^{\KS}_{\per}[\rho](\bfx) :=
  \Biggl[\frac{1}{2\pi \ri}\oint_{\mathscr{C}} \bigl(\lambda - H^{\KS}_{\per}[\rho] \bigr)^{-1} d \lambda \Biggr] (\bfx, \bfx),
\end{equation}
where the contour $\mathscr{C}$ lies in the resolvent set and
separates the occupied and unoccupied spectra.  For the periodic
Kohn-Sham model, we thus recast the problem as a fixed point equation
\begin{equation} \label{eq:fixedPoint_per}
  \rho = \mathcal{F}^{\KS}_{\per}[\rho].
\end{equation}
Using the electron density $\rho$ as the basic variable is more
convenient in studying the scaling limit than the Kohn-Sham orbitals
(Bloch waves), which will be discussed next.

\subsection{Scaling limit for the periodic model}

Starting from the fixed point equation~\eqref{eq:fixedPoint_per}, valid for instance for a periodic configuration $\LL$, we would like to find other solutions when the crystal is deformed.

We consider the electronic structure of an elastically deformed
system in the scaling limit that the lattice parameter goes to $0$. To
setup the atomic configuration, we assume a lattice structure for the
undeformed system. The atoms are located at $\veps \mathbb{L}$, where
$\mathbb{L}$ is a Bravais lattice, and the lattice parameter $\veps$
will serve as the scaling parameter in the limit, which can be understood as the ratio of the lattice parameter and the characteristic length scale of the system.  

For simplicity, we
assume that $\Omega$ coincides with the unit cell of $\mathbb{L}$. Thus, in $\Omega$, atoms are located at
$\Omega \cap \veps \mathbb{L}$. The system consists of $\veps^{-3}$ atoms and
correspondingly $N \veps^{-3}$ electrons where $N$ is the number of
valence electrons per atom.

Fix a smooth function $u: \mathbb{R}^3 \to \mathbb{R}^3$ of the form
$u(\bfx) = B \bfx + u_{\text{per}}(\bfx)$, where $B$ is a $3 \times 3$ matrix
and $u_{\text{per}}$ is periodic with respect to $\Omega$. The
deformed atom locations  are
\begin{equation}
  \bfY_i^{\veps} = \bfX_i^{\veps} + u( \bfX_i^{\veps}), \qquad \bfX_i^{\veps} \in \veps \mathbb{L}.
\end{equation}
Correspondingly the background charge distribution is given by
\begin{equation}
  \mnuceps(\bfy) = \sum_{\bfX_i \in \mathbb{L}} \ma^{\veps}(\bfy - \bfY_i^{\veps}),
\end{equation}
where $\ma^{\veps}$ is the rescaled version of the charge contribution
from each individual atom (recall that the lattice parameter is scaled
to $\veps$):
\begin{equation}
  \ma^{\veps} = \veps^{-3} \ma(\cdot / \veps).
\end{equation}

As the lattice parameter is scaled to be $\veps$, the Kohn-Sham
Hamiltonian needs to be rescaled correspondingly as
\begin{equation}
  H^{\KS}_{\veps, u}[\rho] =  -\frac{\veps^2}{2} \Delta +
  V_{\text{H}}^{\veps}[\rho] + V_{\text{xc}}[\rho],
\end{equation}
where the Hartree potential $V_{\text{H}}^{\veps}$ solves
\begin{equation}
  -\Delta V_{\text{H}}^{\veps} = 4\pi \veps (\rho - \mnuceps).
\end{equation}
Thus the electron density of the deformed system is determined by
the fixed point of the Kohn-Sham map
\begin{equation}\label{eq:deformedKohnShammap}
  \rho(\bfr) = \mathcal{F}^{\KS}_{\veps, u}[\rho](\bfx) :=
  \Biggl[\frac{1}{2\pi \ri}\oint_{\mathscr{C}} \bigl(\lambda - H^{\KS}_{\veps, u}[\rho] \bigr)^{-1} d \lambda \Biggr] (\bfr, \bfr).
\end{equation}

In order to make sense of the Kohn-Sham map defined in
\eqref{eq:kohnshammap} and \eqref{eq:deformedKohnShammap}, we require a gap between the occupied and
unoccupied spectrum, and thus we make the following assumption for the
undeformed system. The gap of the effective Hamiltonian of the perturbed system follows from a perturbation argument. 

\begin{assumption}[Insulating undeformed system]
\label{assump:gap}There exists a
  $\Omega$-periodic $\rho_0 \in C^{\infty}(\mathbb{R}^3)$, that is
  positive and uniformly bounded away from zero, such that
  \begin{itemize}
  \item The spectrum of the Hamiltonian
    $H^{\KS}_{\veps = 1, 0}[\rho_0]$ has a positive gap between the
    occupied and unoccupied spectra.
  \item $\rho_0$ is a fixed point of the Kohn-Sham map:
    \begin{equation}
      \rho_0(\bfr) = \mathcal{F}^{\KS}_{\veps = 1, 0}[\rho_0](\bfr) =
      \Biggl[\frac{1}{2\pi \ri}\oint_{\mathscr{C}} \bigl(\lambda - H^{\KS}_{\veps=1,  0}[\rho_0] \bigr)^{-1} d \lambda \Biggr] (\bfr, \bfr),
    \end{equation}
    where $\mathscr{C}$ is a contour in the resolvent set enclosing
    the occupied spectrum.
  \end{itemize}
\end{assumption}

\subsection{Cauchy-Born rule for electronic structure}

The question of the scaling limit is to characterise the electron
density, as a solution to the Kohn-Sham equation
\eqref{eq:deformedKohnShammap}, when the deformation is elastic in the
sense that the deformation gradient is not too large. This is
motivated by the Cauchy-Born rule for passing from atomistic models to
elastic models, where the analogous question for DFT is to pass from
electronic structure models to continuum elastic models.
The scaling limit for Thomas--Fermi--von Weizsäcker model was studied
by Blanc, Le Bris and Lions in \cite{BlancLeBrisLions:2002}, see Section~\ref{sec:tfw:scaling limit}. 

For the Kohn-Sham type models, the scaling limit was proved by E and
Lu in \cite{ELu:2011, ELu:2013} under the stability conditions on the
level of linear response of the undeformed system. In order to state
the stability assumptions, let us introduce the linearised Kohn-Sham
map for the undeformed system
\begin{equation}
  (\mathcal{L}_0 w) =  \Biggl[\frac{\delta \mathcal{F}^{\KS}_{\veps = 1, 0}[\rho_0]}{\delta \rho}(w) \Biggr].
\end{equation}
It has been established \cite{ELu:2013} that $\mathcal{L}_0$ is a
bounded linear operator on the space
$\mathcal{X}_n := \dot{H}^{-1}_{\per}(n \Omega) \cap H^2_{\per}(n
\Omega)$
for every $n \in \mathbb{N}$, where $H_{\per}^2(n\Omega)$ stands
for the periodic Sobolev space with square integrable second
derivatives and $\dot{H}^{-1}_{\per}(n\Omega)$ is the homogeneous
Sobolev space with index $-1$ on the domain $n \Omega$.

\begin{assumption}[Stability of charge density wave response]
\label{assump:plasmon} For every $n \in \mathbb{N}$, the
  operator $I - \mathcal{L}_0$ is uniformly invertible as an operator
  on $\mathcal{X}_n$.
\end{assumption}

Physically the stability assumption states that the undeformed crystal
is stable with respect to spontaneous charge density wave perturbation
at every scale. In particular, this prevents the possibility of
symmetry breaking as $\veps \to 0$.

We also define the macroscopic permittivity tensor for the undeformed
crystal as
\begin{equation}
  \mathsf{E_0} = \frac12 (\mathsf{A}_0 + \mathsf{A}_0^{\ast}) + \frac{1}{4\pi} \mathsf{I},
\end{equation}
where the $3 \times 3$ matrix $\mathsf{A}_0$ is given by
\begin{equation}
  \mathsf{A}_{0, \alpha\beta} := - 2 \Re \sum_{i}^{\text{occ}} \sum_{a}^{\text{unocc}}
  \int_{\Omega^{\ast}}
  \frac{ \langle u_{a, \bfk}, i \partial_{\bfk_{\alpha}} u_{i, \bfk} \rangle
    \langle u_{a, \bfk}, i \partial_{\bfk_{\beta}} u_{i, \bfk} \rangle^{\ast}}{E_{i, \bfk} - E_{a, \bfk}}
  \frac{d \bfk}{\lvert \Omega^{\ast} \rvert}
  - \bigl\langle g_{\alpha}, \delta_\rho V_{\text{eff}} (I - \mathcal{L}_0)^{-1} g_{\beta} \bigr\rangle
\end{equation}
where
\begin{equation}
  g_{\alpha}(\bfr) := 2 \Re \sum_{i}^{\text{occ}} \sum_{a}^{\text{unocc}}
  \int_{\Omega^{\ast}}
  \frac{ \langle u_{a, \bfk}, i \partial_{\bfk_{\alpha}} u_{i, \bfk} \rangle}{E_{i, \bfk} - E_{a, \bfk}} u_{i, \bfk}^{\ast}(\bfr) u_{a, \bfk}(\bfr)
  \frac{d \bfk}{\lvert \Omega^{\ast} \rvert},
\end{equation}
and $\delta_{\rho} V_{\text{eff}}$ is the linearisation of the
effective potential operator:
$V_{\text{eff}}[\rho] = V_{\text{H}}[\rho] + V_{\text{xc}}[\rho]$ at
$\rho_0$ for the undeformed crystal. The dielectric
permittivity for the reduced Hartree-Fock theory has been studied in
\cite{CancesLewin:2010}.

\begin{assumption}[Stability of dielectric response] 
\label{assump:dielectric} The macroscopic
  permittivity tensor for the undeformed cyrstal $\mathsf{E}_0$ is positive definite.
\end{assumption}

The main result of \cite{ELu:2013} establishes the Cauchy-Born rule for the electronic structure. 

\begin{theorem}\cite[Thm.~5.1]{ELu:2013} 
Under Assumptions~\ref{assump:gap}, \ref{assump:plasmon} and \ref{assump:dielectric},  if the deformation gradient is sufficiently small, then
there exists $\rho^{\veps}_u$ satisfying the Kohn-Sham fixed point
equation \eqref{eq:deformedKohnShammap}, and furthermore, $\rho^{\veps}_u$ can be locally approximated by the
Cauchy-Born rule:
\begin{equation}
  \lVert \rho^{\veps}_u - \veps^{-3} \rho_{\text{CB}}(\bfr/\veps; \nabla u(\bfr)) \lVert_{L^{\infty}} \lesssim \veps^{1/2} \lVert \rho^{\veps}_u \rVert_{L^{\infty}},
\end{equation}
where $\rho_{\text{CB}}(\cdot; A)$ is the electron density of a
homogeneous deformed system with $u(\bfx) = A \bfx$ (which is
well-defined provided $\lvert A \rvert$ is not too large). 
\end{theorem}

The main technical ingredients of the proof of the Theorem is a two-scale analysis of  the linearised Kohn-Sham map. 
As a part of the analysis, the effective potential and the macroscopic dielectric
response of the deformed crystal can be also characterised, we refer the
readers to \cite{ELu:2013} for details.

\bibliographystyle{siam}
\bibliography{reviewtdl}

\begin{thebibliography}{10}

\bibitem{AnantharamanCances:2009}
{\sc A.~Anantharaman and E.~Canc\`es}, {\em Existence of minimizers for
  {K}ohn-{S}ham models in quantum chemistry}, Ann. Inst. Henri Poincar\'e, 26
  (2009), pp.~2425--2455.

\bibitem{Benguria1981}
{\sc R.~Benguria, H.~Brezis, and E.~H. Lieb}, {\em The {Thomas-Fermi-von}
  {W}eizs{\"a}cker theory of atoms and molecules}, Commun. Math. Phys., 79
  (1981), pp.~167--180.

\bibitem{Blanc2006-ej}
{\sc X.~Blanc}, {\em Unique solvability of a system of nonlinear elliptic
  {PDEs} arising in solid state physics}, SIAM J. Math. Anal., 38 (2006),
  pp.~1235--1248.

\bibitem{BlancCances:2005}
{\sc X.~Blanc and E.~Canc{\`e}s}, {\em Nonlinear instability of
  density-independent orbital-free kinetic-energy functionals}, J. Chem. Phys.,
  122 (2005), p.~214106.

\bibitem{BlancLeBrisLions:2002}
{\sc X.~Blanc, C.~Le~Bris, and P.-L. Lions}, {\em From molecular models to
  continuum mechanics}, Arch. Ration. Mech. Anal., 164 (2002), pp.~341--381.

\bibitem{CancesCao2020}
{\sc E.~Canc{\`e}s, L.~Cao, and G.~Stoltz}, {\em Removing a slab from the
  {F}ermi sea: the reduced {H}artree-{F}ock model}, Nonlinearity, 33 (2020),
  pp.~156--195.

\bibitem{Cances2008}
{\sc E.~Canc{\`e}s, A.~Deleurence, and M.~Lewin}, {\em {A new approach to the
  modeling of local defects in crystals: the reduced Hartree-Fock case}},
  Commun. Math. Phys., 281 (2008), pp.~129--177.

\bibitem{Cances2008non}
\leavevmode\vrule height 2pt depth -1.6pt width 23pt, {\em Non-perturbative
  embedding of local defects in crystalline materials}, J. Phys. Condensed
  Matter, 20 (2008), p.~294213.

\bibitem{Cances2011-ot}
{\sc E.~Canc{\`e}s and V.~Ehrlacher}, {\em Local defects are always neutral in
  the {Thomas--Fermi--von W}eisz{\"a}cker theory of crystals}, Arch. Ration.
  Mech. Anal., 202 (2011), pp.~933--973.

\bibitem{Cances2020numerical}
{\sc E.~Canc\`{e}s, V.~Ehrlacher, D.~Gontier, A.~Levitt, and D.~Lombardi}, {\em
  Numerical quadrature in the {B}rillouin zone for periodic {S}chr\"{o}dinger
  operators}, Numer. Math.,  (2020).

\bibitem{LahbabiDisordered}
{\sc E.~Canc{\`e}s, S.~Lahbabi, and M.~Lewin}, {\em Mean-field models for
  disordered crystals}, J. Math. Pures Appl., 100 (2013), pp.~241--274.

\bibitem{CancesLewin:2010}
{\sc E.~Canc\`es and M.~Lewin}, {\em The dielectric permittivity of crystals in
  the reduced {H}artree-{F}ock approximation}, Arch. Ration. Mech. Anal., 1
  (2010), pp.~139--177.

\bibitem{Catto1998}
{\sc I.~Catto, C.~{Le Bris}, and P.-L. Lions}, {\em Sur la limite
  thermodynamique pour des mod{\`e}les de type {H}artree et {H}artree-{F}ock},
  C. R. Acad. Sci. Paris, 327 (1998), pp.~259--266.

\bibitem{Catto1998_book}
\leavevmode\vrule height 2pt depth -1.6pt width 23pt, {\em {The mathematical
  theory of thermodynamic limits: {T}homas-{F}ermi type models}}, Oxford
  University Press, 1998.

\bibitem{Catto2001}
\leavevmode\vrule height 2pt depth -1.6pt width 23pt, {\em {On the
  thermodynamic limit for Hartree-Fock type models}}, Ann. Inst. H.
  Poincar{\'e} (C), 18 (2001), pp.~687--760.

\bibitem{Catto2002}
\leavevmode\vrule height 2pt depth -1.6pt width 23pt, {\em On some periodic
  {H}artree-type models for crystals}, Ann. Inst. H. Poincar{\'e} (C), 19
  (2002), pp.~143--190.

\bibitem{ChenNazarOrtner2019}
{\sc H.~Chen, F.~Q. Nazar, and C.~Ortner}, {\em Geometry equilibration of
  crystalline defects in quantum and atomistic descriptions}, Math. Models
  Meth. Appl. Sc., 29 (2019), pp.~419--492.

\bibitem{Deleurence2008_thesis}
{\sc A.~Deleurence}, {\em Mod{\'e}lisation math{\'e}matique et simulation
  num{\'e}rique de la structure {\'e}lectronique de cristaux en pr{\'e}sence
  des d{\'e}fauts ponctuels}, PhD thesis, Paris Est, 2008.

\bibitem{ELu:2011}
{\sc W.~E and J.~Lu}, {\em The electronic structure of smoothly deformed
  crystals: {W}annier functions and the {C}auchy-{B}orn rule}, Arch. Ration.
  Mech. Anal., 199 (2011), pp.~407--433.

\bibitem{E2012-ay}
{\sc W.~E and J.~Lu}, {\em Stability and the continuum limit of the
  spin-polarized {Thomas-Fermi-Dirac-von W}eizs{\"a}cker model}, Journal of
  Mathematical Physics, 53 (2012), p.~115615.

\bibitem{ELu:2013}
{\sc W.~E and J.~Lu}, {\em The {K}ohn-{S}ham equation for deformed crystals},
  Memoir Amer. Math. Soc., 221 (2013).

\bibitem{Weinan2007-zd}
{\sc W.~E and P.~Ming}, {\em {Cauchy--Born} rule and the stability of
  crystalline solids: Static problems}, Arch. Ration. Mech. Anal., 183 (2007),
  pp.~241--297.

\bibitem{georgii2011gibbs}
{\sc H.-O. Georgii}, {\em Gibbs measures and phase transitions}, vol.~9, Walter
  de Gruyter, 2011.

\bibitem{Gontier:2014}
{\sc D.~Gontier}, {\em Existence of minimizers for {K}ohn-{S}ham within the
  local spin density approximation}, Nonlinearity, 28 (2014), pp.~57--76.

\bibitem{Gontier2016convergence}
{\sc D.~Gontier and S.~Lahbabi}, {\em Convergence rates of supercell
  calculations in the reduced {H}artree-{F}ock model}, ESAIM: Math. Model. Num.
  Anal., 50 (2016), pp.~1403--1424.

\bibitem{Gontier2016supercell}
\leavevmode\vrule height 2pt depth -1.6pt width 23pt, {\em Supercell
  calculations in the reduced {H}artree-{F}ock model for crystals with local
  defects}, Appl. Math. Res. Express,  (2016).

\bibitem{GonLewNaz-21}
{\sc D.~Gontier, M.~Lewin, and F.~Q. Nazar}, {\em The nonlinear
  {S}chr{\"o}dinger equation for orthonormal functions: Existence of ground
  states}, Archive for Rational Mechanics and Analysis, 240 (2021),
  pp.~1203--1254.

\bibitem{LeBris:1993}
{\sc C.~Le~Bris}, {\em Quelques probl\`emes math\'ematiques en chimie quantique
  mol\'eculaire}, PhD thesis, Ecole Polytechnique, 1993.

\bibitem{Leslie1985}
{\sc M.~Leslie and M.~Gillan}, {\em {The energy and elastic dipole tensor of
  defects in ionic crystals calculated by the supercell method}}, J. Phys. C,
  18 (1985), pp.~973--982.

\bibitem{Lieb1997-mo}
{\sc E.~H. Lieb}, {\em Thomas-{F}ermi and related theories of atoms and
  molecules}, in The Stability of Matter: From Atoms to Stars: Selecta of
  Elliott H. Lieb, W.~Thirring, ed., Springer Berlin Heidelberg, Berlin,
  Heidelberg, 1997, pp.~259--297.

\bibitem{LiebSimon1977}
{\sc E.~H. Lieb and B.~Simon}, {\em The {T}homas-{F}ermi theory of atoms,
  molecules and solids}, Adv. Math., 23 (1977), pp.~22--116.

\bibitem{Lu2015-rs}
{\sc J.~Lu, V.~Moroz, and C.~B. Muratov}, {\em {Orbital-Free} density
  functional theory of {Out-of-Plane} charge screening in graphene}, J.
  Nonlinear Sci., 25 (2015), pp.~1391--1430.

\bibitem{Makov1995}
{\sc G.~Makov and M.~Payne}, {\em Periodic boundary conditions in \textit{ab
  initio} calculations}, Phys. Rev. B, 51 (1995), pp.~4014--4022.

\bibitem{Monkhorst1976}
{\sc H.~Monkhorst and J.~Pack}, {\em {Special points for {B}rillouin-zone
  integrations}}, Phys. Rev. B, 13 (1976), pp.~5188--5192.

\bibitem{2015-tfw}
{\sc F.~Q. Nazar and C.~Ortner}, {\em Locality of the thomas-fermi-von
  {W}eizs{\"a}cker equations}, Arch. Ration. Mech. Anal., 224 (2017).

\bibitem{Ortner2013-en}
{\sc C.~Ortner and F.~Theil}, {\em Justification of the {Cauchy--Born}
  approximation of elastodynamics}, Arch. Ration. Mech. Anal., 207 (2013),
  pp.~1025--1073.

\bibitem{ReedSimon4}
{\sc M.~Reed and B.~Simon}, {\em {Methods of Modern Mathematical Physics.
  Analysis of Operators}}, vol.~IV, Academic Press, 1978.

\bibitem{Ricaud2018}
{\sc J.~Ricaud}, {\em Symmetry breaking in the periodic
  {T}homas--{F}ermi--{D}irac--von {W}eizs{\"a}cker model}, Ann. H.
  Poincar{\'e}, 19 (2018), pp.~3129--3177.

\bibitem{ruelle1999statistical}
{\sc D.~Ruelle}, {\em Statistical mechanics: Rigorous results}, World
  Scientific, 1999.

\bibitem{Solovej1990-aa}
{\sc J.~P. Solovej}, {\em Universality in the {Thomas-Fermi-von}
  {W}eizs{\"a}cker model of atoms and molecules}, Commun. Math. Phys., 129
  (1990), pp.~561--598.

\bibitem{solovej1991proof}
\leavevmode\vrule height 2pt depth -1.6pt width 23pt, {\em Proof of the
  ionization conjecture in a reduced {H}artree-{F}ock model}, Invent. Math.,
  104 (1991), pp.~291--311.

\bibitem{Von_Weizsacker1935-nj}
{\sc C.~F. Von~Weizs{\"a}cker}, {\em Zur theorie de kernmassen}, Z. Phys., 96
  (1935), p.~431.

\bibitem{WangTeter:1992}
{\sc L.-W. Wang and P.~Teter, Michael}, {\em Kinetic-energy functional of the
  electron density}, Phys. Rev. B, 45 (1992), p.~13196.

\bibitem{WangGovindCarter:1999}
{\sc Y.~A. Wang, N.~Govind, and E.~A. Carter}, {\em Orbital-free kinetic-energy
  density functionals with a density-dependent kernel}, Phys. Rev. B, 60
  (1999), p.~16350.

\end{thebibliography}

\end{document}